\documentclass[preprint]{aastex}

\slugcomment{To appear in the ApJ}

\shortauthors{Worrall \& Birkinshaw}
\shorttitle{X-ray Properties of Radio Galaxies}

\begin{document}

\title{The X-ray emission of 3C~346 and its environment}

\author{D.M.~Worrall\altaffilmark{1} and
M.~Birkinshaw\altaffilmark{1}}

\affil{Department of Physics, University of Bristol, U.K.}

\altaffiltext{1}{Harvard-Smithsonian Center for Astrophysics,
Cambridge, MA 02138}

\email{d.worrall@bristol.ac.uk}

\begin{abstract}

We present a detailed spectral and spatial analysis of the X-ray
properties of the compact and unusual radio galaxy
3C~346, combining information from ROSAT and ASCA.  The
dominant component of X-ray emission ($\sim 10^{44}$ ergs s$^{-1}$ in
each of the 0.5-3 keV and 2-10~keV bands) is unresolved and not
heavily absorbed (intrinsic N$_{\rm H} \lesssim 2 \times 10^{21}$
cm$^{-2}$), with evidence for variability of $32\pm 13$\% over 18
months.  We relate the X-ray emission to radio structures on both
milliarcsecond scales and the arcsecond scales which {\it Chandra\/}
can resolve.  The absence of X-ray absorption, and the
radio/optical/X-ray colors, when combined with previous radio evidence
that the source is a foreshortened FRII, suggest that the radio jets
are seen at an angle to the line of sight of about 30$^\circ$,
intermediate between the radio-galaxy and quasar classes.  Roughly a
third of the soft X-ray emission is from a cluster atmosphere, for
which we measure a temperature of $1.9^{+1.3}_{-0.7}$ keV, making this
the second low-redshift ($z < 0.2$) powerful radio galaxy, after
Cygnus~A, with a measured cluster temperature.  At a jet angle of $\sim
30^\circ$, all the radio structures lie within the core radius of the
cluster, for which the cooling time is sufficiently long that there is
no reason to expect the presence of a cooling flow.  The radio lobes
of 3C~346 are roughly in pressure balance with the external medium
under the assumptions that the energy densities in the magnetic field
and radiating particles balance and that a source of excess pressure
in the radio lobes, commonly invoked in other radio galaxies, is
absent here.

\end{abstract}

\keywords{galaxies:active --- galaxies:clusters:individual (3C 346)
--- galaxies:individual (3C 346) --- galaxies:jets --- radiation
mechanisms:nonthermal --- X-rays:galaxies}

\section{Introduction}\label{sec-intro}

3C~346 is a well-known compact radio source in a 17th magnitude galaxy
at $z=0.161$ \citep[e.g.][]{lrl83}.  Although originally classified as
a member of the physically small class of Compact Steep Spectrum (CSS)
radio sources \citep{fanti85}, \citet{spence91} argue that its
luminous core and small, one-sided (eastwards), distorted jet
structure are best explained by the foreshortening of a normal radio
galaxy by a combination of relativistic beaming and small angle to
the line of sight.  The radio morphology then suggests it would
be a member of the FRII class, as defined by \citet{fr74}, and this is
consistent with its 178~MHz power of $10^{26}$ W Hz$^{-1}$ sr$^{-1}$,
about a factor of five higher than the fuzzy boundary between sources
exhibiting FRI structure and those classed as FRIIs.  Subsequent
authors have generally supported the foreshortened FRII picture, and
in particular \citet{cotton95} use the VLBI radio-core dominance and
jet-to-counterjet ratio to infer an angle to the line of sight of
$\theta < 32^\circ$, and a speed relative to that of light of $\beta >
0.8$.  Unified Models \citep[e.g.][]{bart89} class FRII radio sources
with their jets this close to the line of sight as quasars in contrast
to radio galaxies, and so the absence of an expected strong broad emission-line
region (BELR) needs explanation. \citet{dey94} combine their upper
limit for the flux of H$\beta$ with the emission-line flux of
H$\alpha$ + [N~II] from narrow-band imaging data of \citet{baum88} to
suggest that the nuclear regions are seen through a large extinction
($A_v > 8$~mag), such that the BELR may only be detectable in the
infrared.  However, such a large extinction would then be unusual for
a source of 3C~346's presumed orientation.

At high radio frequencies (15~GHz), 3C~346's one sided wiggly jet
breaks into a series of bright knots which \citet{vanb92} label as B,
C, D, and E with increasing distance from the core.  The brightest
knot, C, at $\sim 3''$ east of the radio core may reside outside the
optical galaxy if the jet is indeed at a small angle to the line of
site.  \citet{dey94} detected excess emission in ground-based $U$ and
red images which they attributed to knot C and its nearby companion,
knot B.  More recently an HST snapshot image with the WFPC2 has
provided a spectacular view of the jet at optical wavelengths (F702W
filter: $\sim 6000-8000$ \AA), with a one-to-one correspondence
between optical and radio features in the jet, including knots B, C,
and D \citep{dekoff96, devries97}.

3C~346 was detected in the X-ray with the {\it Einstein\/} IPC, yielding
$136\pm 13$ counts in 2.9 ks, and giving a 0.5-3~keV luminosity of
$1.4 \times 10^{44}$ ergs s$^{-1}$ to within an estimated $\sim 30\%$
uncertainty \citep{fab84}.  Subsequent longer observations were made
with the ROSAT PSPC, which permitted \citet{hard99} to show that the
X-ray emission is a composite of resolved and unresolved components,
and with ASCA, from which \citet{sam99} derived a power-law spectral
index for the overall X-ray emission as part of a statistical study of
the X-ray emission from a number of radio-loud active galaxies.  In
this paper we present a detailed analysis of the X-ray properties,
combining information from ROSAT and ASCA (\S \ref{sec-results}).  We
discuss the pressure and cooling time of the cluster gas which gives
rise to the extended X-ray emission (\S \ref{sec-cluster}), and we
discuss the likely origin of the unresolved X-ray emission in the
context of what is known from radio and optical measurements (\S
\ref{sec-agn}).  Our conclusions are in \S \ref{sec-conclusions}.

Throughout the paper we adopt a Friedmann cosmological model with
$H_o$ = 50 km s$^{-1}$ Mpc$^{-1}$, $q_o$ = 0.  At the redshift of
3C~346, 10 arcsec corresponds to 37.5~kpc.

\section{X-ray Analysis and Results}\label{sec-results}

Dates and exposure times for the ROSAT and ASCA X-ray observations are
given in Table~\ref{obs-tab}. Our analysis made use of the IRAF/Post
Reduction Off-line Software (PROS) for ROSAT and FTOOLS and XSPEC for
ASCA.  The ASCA data are from the two Gas Imaging Spectrometers (GIS)
and the two Solid-State Spectrometers (SIS).  The SIS data were taken
in a mixture of 2-CCD faint and bright mode; the faint-mode data were
converted to bright mode by the standard processing software before
analysis.  Our screening used standard recommended procedures, and
values adopted included an elevation angle $> 5$ degrees for dark
earth (and 12 degrees for bright earth with the SIS), and a cut-off
rigidity of 6 GeV/c for the SIS and 4 GeV/c, with extra screening, for
the GIS.  Four 8-sec intervals of anomalously high background counts
in the SIS1 were excluded from the data.  Our spatial analysis of the
ROSAT data used generalized software \citep{birk94, worr94} for
fitting radial profiles to combinations of models convolved with the
instrument Point Response Function (PRF).

\subsection{ROSAT Spatial and Spectral Results}

\begin{figure}
\epsscale{0.5}
\plotone{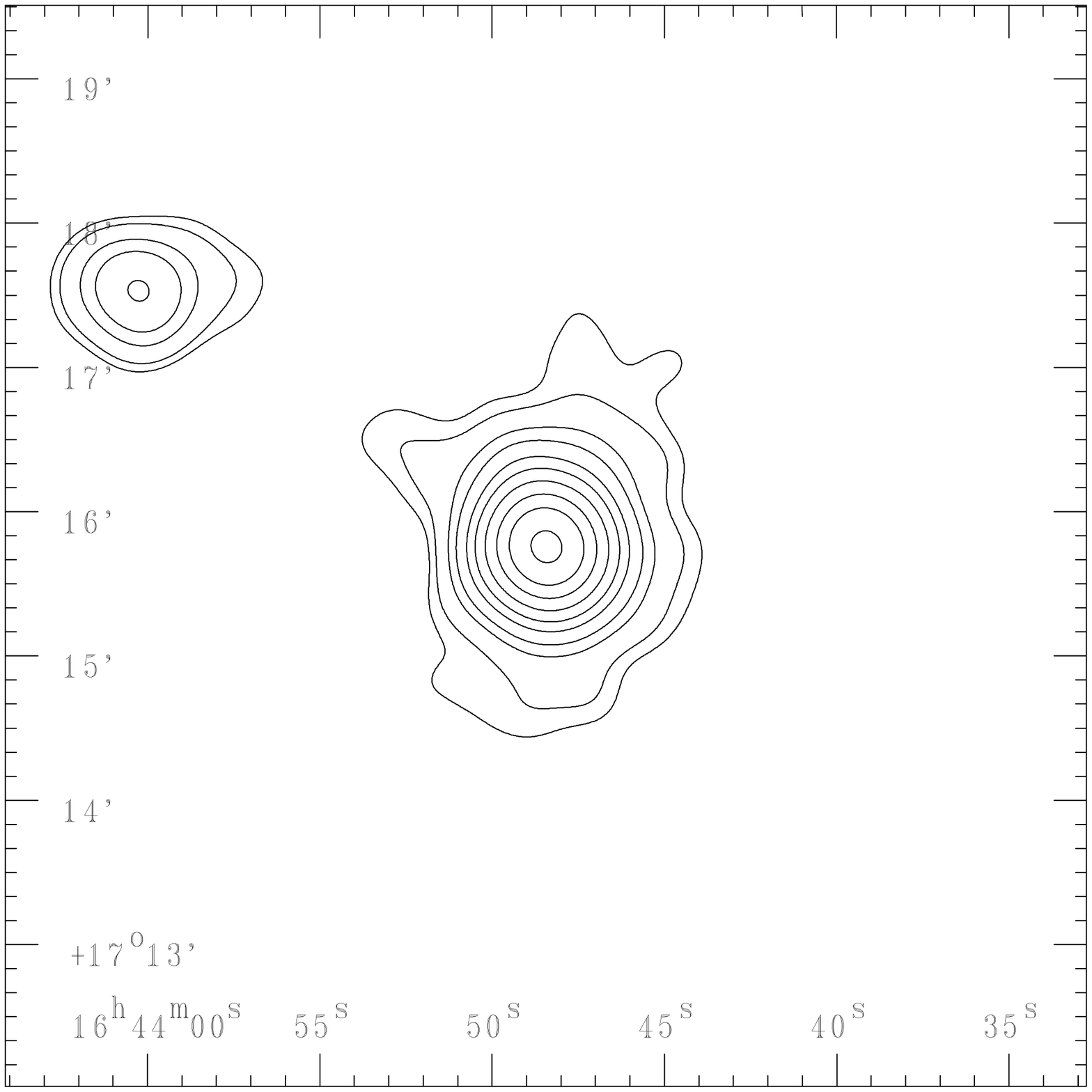}
\caption{ROSAT PSPC image centered on 3C 346, in J2000 equatorial
coordinates.  Data have been smoothed with a Gaussian of $\sigma =
12$~arcsec.  Pixels are $0.5 \times 0.5$ arcsec square, and the lowest
contour corresponds to $3\sigma$ significance.  Contour levels, in
cts/pixel, are 0.003, 0.004, 0.007, 0.01, 0.016, 0.023, 0.034, 0.05,
0.07 and 0.1.  3C~346's radio and optical jet structures span an
angular size less than 4 arcsec (see text), and lie well within the
PRF of the ROSAT PSPC, as does the combined optical extent of 3C~346's
host galaxy and its nearby companion \citep{dekoff96}.
Lower-frequency (1.5 GHz) radio emission extends over a region $\sim
14'' \times 12''$ \citep{vanb92}, also within the X-ray PRF.
\label{cont346}}
\end{figure}

The ROSAT PSPC image centered on 3C~346 is shown as a contour plot in
in Figure~\ref{cont346}.  A re-registration of the X-ray image by
5~arcsec (within ROSAT's absolute positional accuracy) aligns the
centroid of the X-ray emission with the radio core of 3C~346, and the
X-ray source 3.34 arcmin to 3C~346's northeast with a star of
late-type color index listed in the USNO catalog (R=12.8 mag, B=15.2
mag).  The northeast source, at an X-ray flux of $\sim 6 \times
10^{-14}$ ergs cm$^{-2}$ s$^{-1}$ (0.2 - 2 keV), has an
X-ray-to-optical flux ratio consistent with the range measured for
late-type stars in the EMSS survey \citep[e.g.][]{scio95}, and,
although there are insufficient counts to constrain well the X-ray
spectrum, it can be fitted with gas components of temperature $\sim 2
\times 10^6$ and $\sim 10^7$ K, consistent with findings for late-type
stellar coronae \citep{prei97}.

\begin{figure}
\epsscale{1.0}
\plottwo{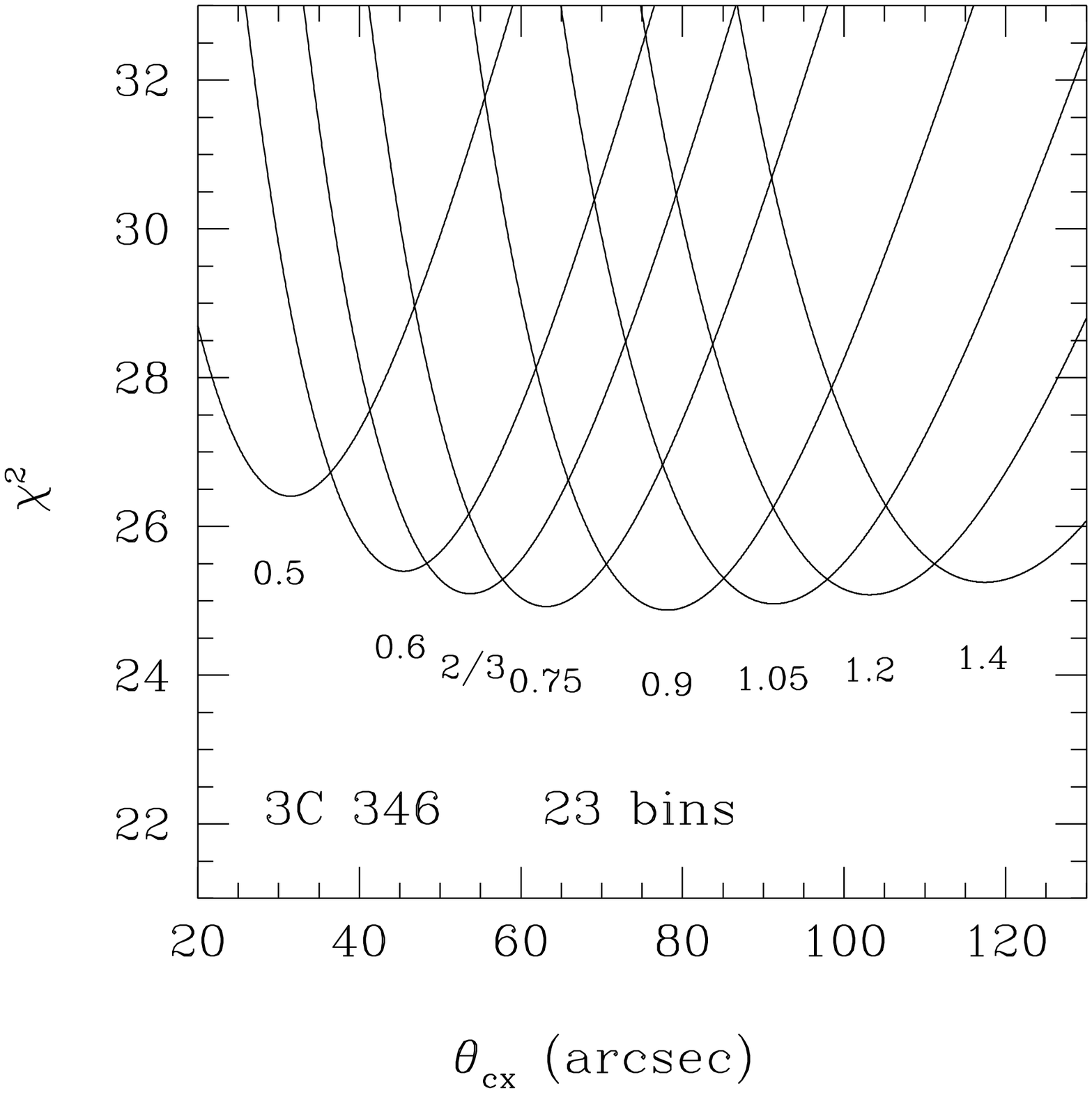}{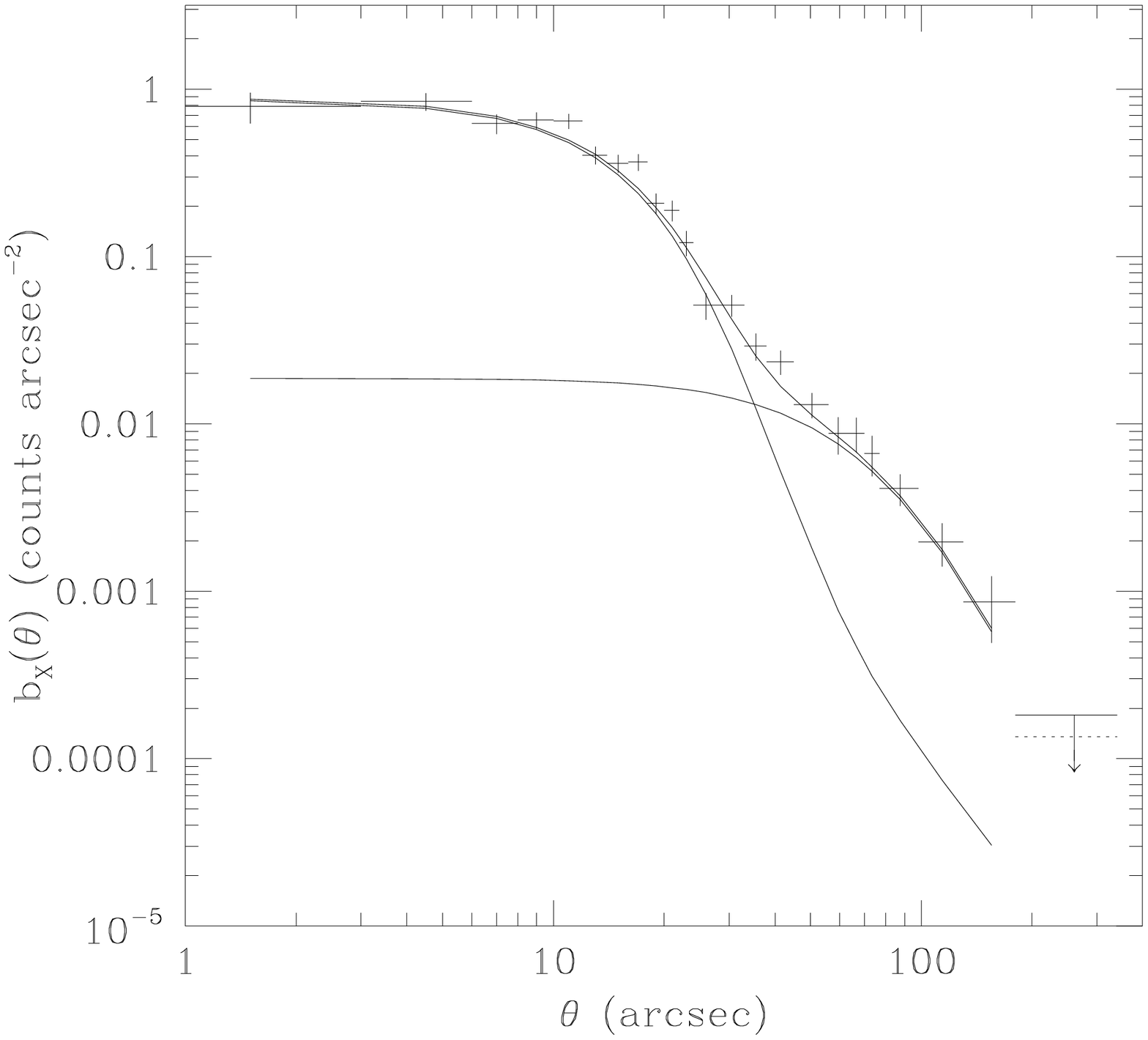}
\caption{Result of fitting a $\beta$-model plus point source to the
PSPC radial profile of 3C~346.  Left: -- $\chi^2$ versus core radius
of the $\beta$ model, for different values of $\beta$.  The fits are
relatively insensitive to the value of $\beta$ which is highly
correlated with core radius, $\theta_{\rm cx}$.  Right: - data show
the background-subtracted radial profile, and solid curves the two
model components of the best-fit model.  The $\beta$-model (broader
solid curve) is for $\beta = 0.9$, $\theta_{cx} =78''$. The narrow
solid curve is the PSPC PRF.  The dotted line shows the contribution
of the model to the background region, as taken into account in the
fitting.
\label{beta346}}
\end{figure}

Figure~\ref{beta346} shows the X-ray radial profile that we extracted
for 3C~346, excluding a circle of radius $2'$ around the source to the
northeast.  The boundaries of the 23 bins were selected so as to give
at least 20 counts per bin, and only counts in the energy band
0.2-1.9~keV, where the PRF is well modelled, were used.  Background
was measured from an annulus of radii 3 and 5.7 arcmin, again
excluding the nearby source.  A fit of this radial profile to the PRF
gives an unacceptable $\chi^2$ of 208, indicating the presence of
extended X-ray emission, whereas the X-ray source to the northeast is
consistent with being point-like.  A single-component
$\beta$-model\footnote{Counts per unit area per unit time at radius
$\theta$ proportional to $(1 + {\theta^2 \over \theta_{\rm
cx}^2})^{0.5 - 3 \beta}$.}, a description of gas in hydrostatic
equilibrium \citep[e.g.][]{sar86}, also gives a poor fit to the
emission centered on 3C~346, with $\chi^2_{\rm min} \sim 40$ for
$\beta \sim 0.5$ and an unacceptably small core radius of $< 0.05$
arcsec.  The composite of a point source and $\beta$-model gives
acceptable fits which are relatively insensitive to the value of
$\beta$, which is itself highly correlated with core radius,
$\theta_{\rm cx}$ (Fig.~\ref{beta346}).  In \citet{hard99} we present
results for one set of acceptable parameters.  Here we extend that
work by exploring the uncertainties in physical parameters resulting
from the full range in spatial and spectral model parameters.

\begin{figure}
\epsscale{1.0}
\plottwo{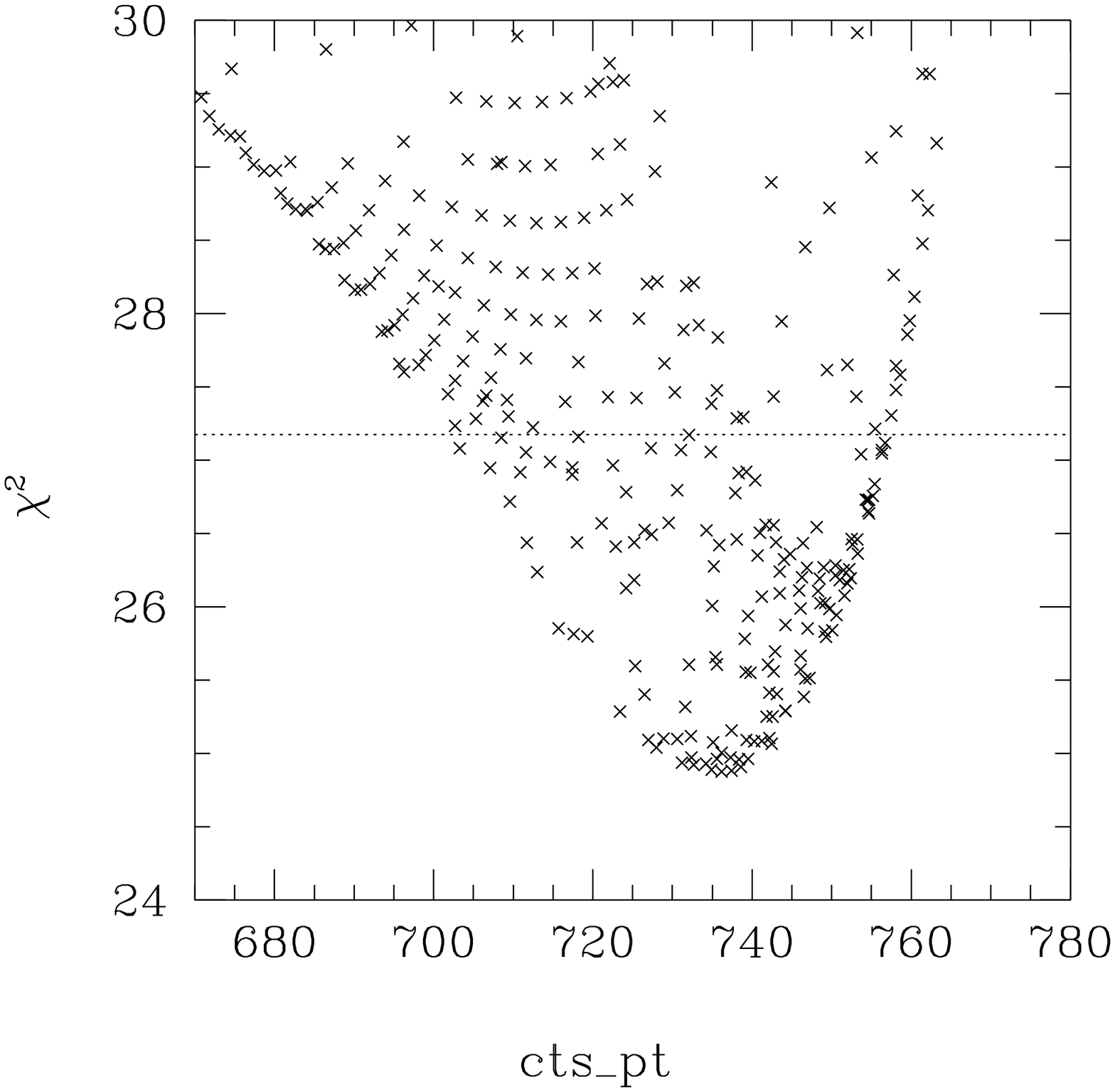}{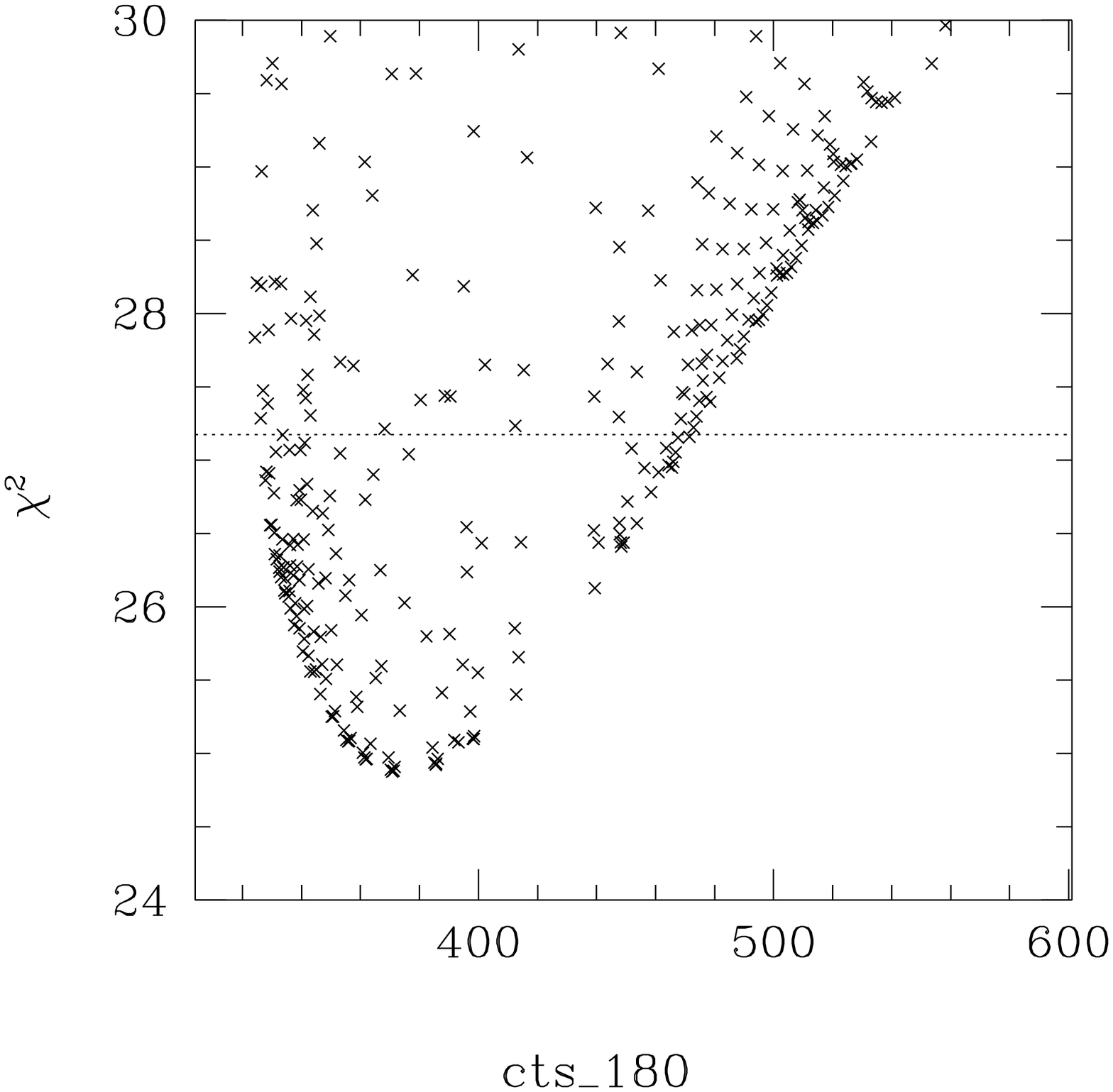}
\caption{ROSAT PSPC counts in the
unresolved component, cts\_pt (Left) and
the $\beta$ model (out to a radius of $3'$
and including the $\sim 18\%$ correction
for missing counts from the region of the nearby source to the
northeast, cts\_180: Right) from
the two-component spatial analysis.
In each plot, a cross marks the position for a fit of chosen
$\beta$ and $\theta_{\rm cx}$ (where the normalizations of the two
model components were free
parameters of the fit), and necessarily large ranges of
$\beta$ and $\theta_{\rm cx}$ were sampled
to provide full ranges in
cts\_pt and cts\_180 for all values of $\chi^2$ shown.  The dotted lines are at
$\chi^2_{\rm min} + 2.3$, corresponding to $1\sigma$ for two interesting
parameters.
\label{radialcts}}
\end{figure}

\begin{figure}
\epsscale{0.5}
\plotone{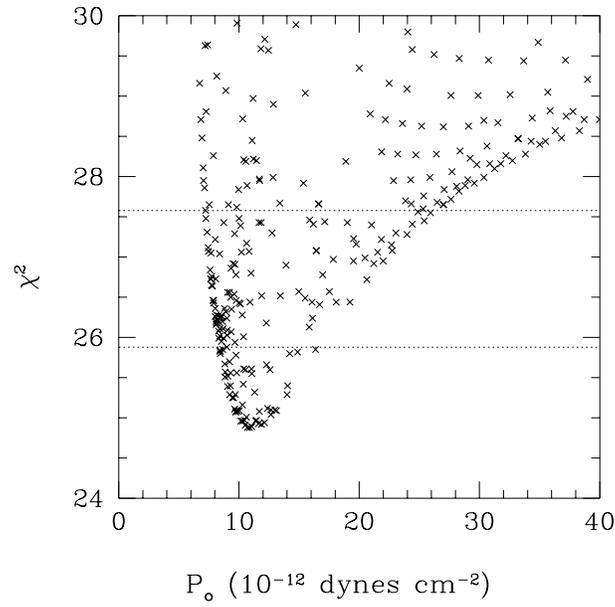}
\caption{Similar to Figure~\ref{radialcts} but for
central gas pressure. Dotted lines are at
$\chi^2_{\rm min} + 1$ and $\chi^2_{\rm min} + 2.7$,
corresponding to $1\sigma$ and 90\% confidence, respectively, for
one interesting parameter.
An additional error in central gas pressure contributes to the value
in Table~\ref{cluster-tab} due to the
uncertainty in temperature.
\label{radialpressure}}
\end{figure}

Figure~\ref{radialcts} shows $\chi^2$ as a function of counts
in the unresolved component
and in the $\beta$ model (out to a radius of $3'$),
where each cross shows the minimum value of $\chi^2$ for
a given trial combination of $\beta$ and $\theta_{\rm cx}$, where the model
normalizations are free parameters of the fit.  Fits are performed
over a broader range of $\beta$ and $\theta_{\rm cx}$ than shown in
the left-hand plot of Figure~\ref{radialcts}, such that all
combinations of parameters giving a particular value of $\chi^2$ (less
than $\chi^2_{\rm min} + 5$) are adequately sampled.  The model being
fitted to the data has four parameters (point-source normalization,
beta-model normalization, $\beta$, and $\theta_{\rm cx}$), and in
principle could be written in such a way as to make any quantity
which is dependent on one or more of these four parameters to be itself a
parameter of the model.  In practice, the trade-off between
computing time and programming time makes it more efficient to use
the brute-force method of computing $\chi^2$ for many parameter-value
combinations and plotting $\chi^2$ as a function of the
quantity of interest, as shown in Figures~\ref{radialcts} and
\ref{radialpressure}.

The counts in the point-source model are highly correlated with the
counts in the $\beta$ model and so, since we are interested in
comparing the division of counts between the two components found in
this spatial analysis with
spectroscopic results, our $1\sigma$ errors are given by $\chi^2_{\rm
min} + 2.3$ ($1\sigma$ for two interesting parameters).  Our spatial
fitting thus finds $736^{+21}_{-34}$ ROSAT PSPC counts (0.2-1.9~keV)
in the unresolved component, and $371^{+104}_{-46}$ in resolved
emission out to a radius of $3'$ (including the $\sim 18\%$ correction
for missing counts from the region of the nearby source to the
northeast).  The resolved emission is well concentrated within a
radius of $3'$, as indicated by the shape of the radial profile and
the preference for values of $\beta$ close to unity
(Fig.~\ref{beta346}); extending the radius to $4'$ increases the
counts in the $\beta$ model only by 5.6\%.

Spectral fits to the ROSAT PSPC data (over the same spatial on-source
and background regions as used in our spatial analysis, but over a
broader energy range) give acceptable fits to either a power law of
energy index 0.75 ($\chi^2_{\rm min} = 18.2$ for 23 degrees of
freedom) or a Raymond-Smith thermal model with 30\% solar abundances
and $kT = 2.9$~keV ($\chi^2_{\rm min} = 18.3$).  In neither case is
excess absorption over the column density through the Galaxy required,
with $3\sigma$ upper limits of $1.7 \times 10^{21}$ cm$^{-2}$ and $6
\times 10^{20}$ cm$^{-2}$ for the power-law and thermal models,
respectively.  Interestingly, while a combination of power-law and
thermal spectral components does not decrease $\chi^2$ sufficiently to
argue, using an F-test, that two components are required based on the
spectral analysis alone,
the two-component
fit settles on parameters which give roughly the same division of
counts between the power-law and thermal components as did the spatial
analysis between the unresolved and resolved components, respectively.

The extended X-ray emission is undoubtedly dominated by thermal
radiation from hot gas.  3C~346's radio lobes lie within $10''$ of the
core, and so any upscattering of photons by the relativistic plasma
occurs well within the PSPC's PRF (see profile in Fig.~\ref{beta346}),
and cannot account for the resolved component.  The fact that the
extended emission is well fitted by a $\beta$-model, and the {\it
post-facto\/} result that the luminosity and temperature of this
component fall in the range measured for the atmospheres of other
radio galaxies, further support a thermal origin for these X-rays.

We model the unresolved emission with a power-law spectrum, on the
assumption that any contributions to this component from discrete
sources and galaxy-scale gas are negligible.  That this is
overwhelmingly likely is clear because the soft X-ray luminosity of
the unresolved emission is ten times higher than that of any
elliptical galaxy in the compilation of \citet{fab92}, and a thousand
times higher than for typical elliptical galaxies.  Furthermore, if we
assume only 30\% of the unresolved emission is thermal, and model it
as a sphere of radius $11.5''$, half the FWHM of the PSPC PRF, we find
that the emitting gas would have a cooling time much less than the
Hubble time, and a mass deposition rate of $\sim 90 M_\odot$/yr. Such
a cluster-scale mass deposition in what would be a galaxy-scale
cooling flow is unreasonable.

Since the extracted spectrum should contain all the counts from the
unresolved component,
we performed two-component spectral fits with the constraint
that the power-law component provide between 702 and 757
counts between 0.2 and 1.9~keV.  For these fits, the hydrogen column
density was fixed at the Galactic value (since single-component fits
required no excess absorption), and abundances were fixed at 30\%, the
typical value for cluster gas \citep{arnaud92}.  The fits
constrained the temperature of the gas to be $kT =
1.9^{+1.3}_{-0.7}$~keV and the power-law energy index, $\alpha$,
($f_\nu \propto \nu^{-\alpha}$) to be $0.69^{+0.16}_{-0.14}$ (other
temperature and slope combinations being inconsistent with the
required counts in the power-law model).  
The statistical uncertainty in the division of counts between the two
components dominates the errors in the spectral model parameters.
For example, allowing the abundance percentage
to vary by a factor of three from the adopted value
(i.e. allowing it to cover a range from 10\% to 90\%)
increases the lower and upper
uncertainties in temperature by at most 20\% and 40\%, respectively.
Tables~\ref{cluster-tab} and
\ref{agn-tab} summarize the results for the cluster gas and unresolved
(power-law) emission (which we attribute to the active galaxy).  The
tables include the luminosities in various spectral bands, where the
uncertainties are dominated by the errors in counts represented in
Figure~\ref{radialcts}.

Now that we have measured a gas temperature, we return to the spatial
fits to compute other parameters of interest (Table~\ref{cluster-tab})
such as the central pressure and density, and their equivalents at a
radius of $10''$, roughly the projected outer radius of the radio
lobes, which appear as an elliptical halo around the resolved jet
features \citep{vanb92}.  Figure~\ref{radialpressure} is a similar
plot to Figure~\ref{radialcts}, but for central gas pressure, assuming
a temperature of $kT = 1.9$~keV.  For $kT$ between 1.2 and 3.2 keV,
the PSPC count-rate per unit emission measure is independent of
temperature, to within the statistical uncertainties.  This means that
our density measurements are unaffected by the temperature
uncertainty, whereas the pressure and cooling-time errors
given in Table~\ref{cluster-tab} have been increased to include
a contribution from the temperature range, since pressure is
$\propto kT$ and cooling time is approximately $\propto$ $\sqrt{kT}$.

\subsection{ASCA Spectral Results using ROSAT Spatial Analysis}

ROSAT has found that the soft X-ray emission from 3C~346 divides
roughly as two-thirds point-like and one-third extended emission
concentrated within a radius of $\sim 3'$.  Visual inspection of the
ASCA images showed no evidence for resolved X-ray
emission, consistent with ASCA's larger PRF as compared with ROSAT.
For the results presented here, we extracted spectra from circular
regions centered on the source position, using a radius of $4'$ for
the SIS, and the larger value of $6'$ for the GIS to accommodate this
detector's contribution to the PRF.  Background was measured from
off-source regions on the images.  Although the location of the X-ray
emitting star to the northeast of 3C~346 (Fig.~\ref{cont346}) falls
within our on-source regions, the X-ray emission is weak and soft
compared with that from 3C~346, and contamination in ASCA's somewhat
harder spectral band is negligible.  Spectra were
rebinned to a minimum of 20 counts per bin before we performed
$\chi^2$ fitting of the SIS and GIS data jointly to various models.
We followed the recommended procedure of excluding energies below
0.6~keV (0.4~keV for the SIS) and above 10~keV, where the detector
spectral responses are uncertain.

\begin{figure}
\epsscale{0.7}
\plotone{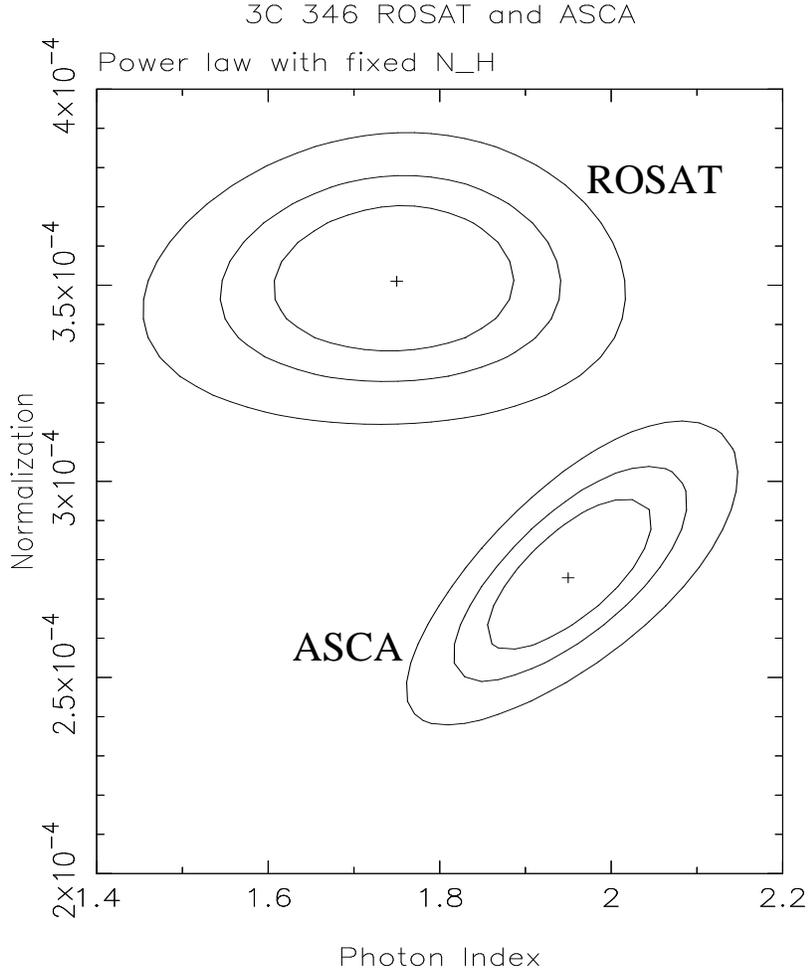}
\caption{Uncertainties in normalization and power-law photon index
($\alpha + 1$) for single-component power-law spectral
fits to the ROSAT PSPC (upper contours) and ASCA combined SIS and GIS
data (lower contours).  The contours are at
$\chi^2_{\rm min}$ + 2.3, 4.61 and 9.21, corresponding to confidence
levels of $1\sigma$, 90\% and 99\% for two interesting parameters.
The contours do not overlap, suggesting a combination of
source variability and the model being inadequate to describe
the data, which are known from the spatial analysis
to include thermal emission.
\label{ascapspc}}
\end{figure}

As is the case for ROSAT, the data give an acceptable fit to a
single-component power law with no excess absorption over the
line-of-sight column density through our galaxy ($\chi^2$ = 159 for
154 degrees of freedom).  Uncertainty contours for power-law photon
index ($\alpha$ + 1) and normalization are shown in
Figure~\ref{ascapspc}.  Our spectral-index measurement of $1.95 \pm 0.11$
($\chi^2_{\rm min} + 2.7$), upon which the GIS and SIS separately
agree well, is a little steeper than that found independently by
\citet{sam99} from the same observation ($1.81 \pm 0.12$), but results
agree within the uncertainties.
Excess intrinsic absorption is less well constrained than with ROSAT,
and the 3~$\sigma$ upper limit on line of sight hydrogen column
density is $2.9 \times 10^{21}$ cm$^{-2}$.

Although we know from the spatial analysis of the ROSAT data that a
single component power law is an inadequate description of the X-ray
emission from 3C 346, is is instructive to compare the ROSAT and ASCA
results for this spectral model.  Figure~\ref{ascapspc} provides the
first indication that the X-ray emission may have varied between the
ROSAT and ASCA observations.

\begin{figure}
\epsscale{0.7}
\plotone{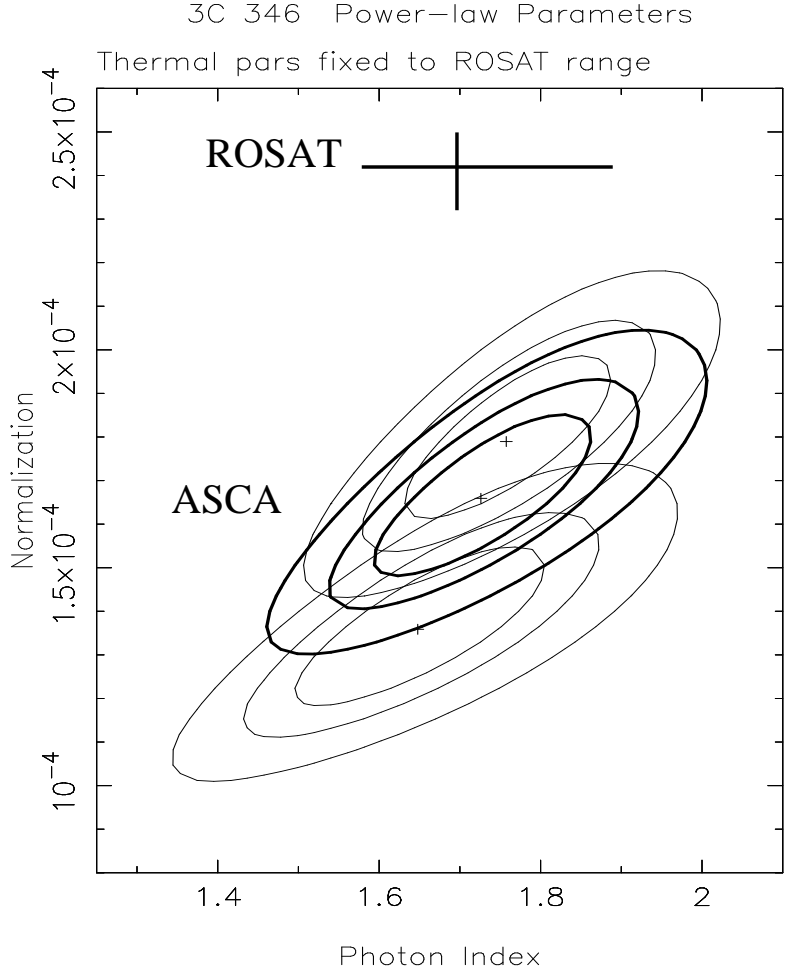}
\caption{Uncertainties in power-law normalization and photon index
($\alpha + 1$) for two-component (thermal plus power-law) spectral
fits to the ASCA combined SIS and GIS data.  The contours are at
$\chi^2_{\rm min}$ + 2.3, 4.61 and 9.21, corresponding to confidence
levels of $1\sigma$, 90\% and 99\% for two interesting parameters.
The set of heavy contours are where the thermal spectrum is fixed to
the best-fit parameters from the ROSAT analysis.  The two other sets
of contours correspond to the upper and lower bounds on the thermal
component as deduced from ROSAT, taking into account both statistical
and modelling uncertainties.  The normalization is in units of
photons cm$^{-2}$ s$^{-1}$ keV$^{-1}$ at 1~keV: a normalization of
$10^{-4}$ corresponds to 0.0663$\mu$Jy.  The ROSAT analysis finds a
photon index of $1.69^{+0.16}_{-0.14}$ 
 and a normalization between $2.33 \times
10^{-4}$ and $2.5 \times 10^{-4}$ ($1\sigma$ for two interesting
parameters).  This suggests that the power-law component decreased in
intensity by $\sim 30\%$ between the ROSAT and ASCA observations,
while staying roughly constant in spectral slope.
\label{ascaconts}}
\end{figure}

The statistical errors on the ASCA data are too large to constrain
well the parameters for a two-component, power law and thermal, fit.
We have therefore fixed the thermal component to be within the
temperature and normalization bounds determined using the ROSAT data.
Figure~\ref{ascaconts} compares the ASCA-derived uncertainties in
power-law normalization and photon spectral index with the range found
from ROSAT.  There is no line of fixed normalization in
Figure~\ref{ascaconts} which goes through both the ROSAT and ASCA
error regions, even for different photon spectral indices -- a case
which would have been interpreted as a change in spectral slope
between the ROSAT and ASCA energy bands.  Instead, we conclude that
the power-law component has decreased by $\sim 32 \pm 13$\% between
the ROSAT and ASCA measurements.

The only other X-ray measurement of which we are aware is that with
the {\it Einstein\/} Observatory reported by \citet{fab84}.  Our
estimates of the total 0.5-3 keV luminosity from ROSAT and ASCA
(Tables~\ref{cluster-tab} and \ref{agn-tab}) of $1.4^{+0.4}_{-0.2}
\times 10^{44}$ ergs s$^{-1}$ and $1.1^{+0.4}_{-0.3} \times 10^{44}$
ergs s$^{-1}$, respectively, are both consistent with the {\it
Einstein\/} luminosity of $1.4 \pm 0.4 \times 10^{44}$ ergs s$^{-1}$.

\section{The Cluster Gas}\label{sec-cluster}

Table~\ref{cluster-tab} summarizes the properties of the X-ray
emitting gas.  3C~346 is the second low-redshift ($z < 0.2$) powerful
radio galaxy, after Cygnus~A (3C~405), with a measured cluster
temperature.  Indeed, the X-ray atmospheres around most powerful radio
galaxies at low redshift are currently undetected, although for
ROSAT-observed sources there is some evidence that gas may well be
present at levels close to the X-ray upper limits \citep{hard00b}.
The cluster around Cygnus~A is considerably hotter and richer than
that around 3C~346, whose atmosphere is more similar to that around a
typical low-power (FRI) radio galaxy (Fig.~\ref{templum}).  The
cooling time of 3C~346's cluster atmosphere (Table~\ref{cluster-tab})
is sufficiently long that there is no reason to invoke the presence of
a cooling flow.  Optical evidence for cluster membership is presented
by \citet{zirb97}, who classifies the cluster as Bautz-Morgan class I,
with a richness of $10\pm 4.6$.

The southwest radio lobe, modelled as lying at an angular distance
between $0''$ and $10''$ from the core, is estimated to have a minimum
pressure in magnetic field and radiating particles of $10^{-11}$ dynes
cm$^{-2}$ \citep{hard00b}.  The closeness of this value to the gas
pressure (Table~\ref{cluster-tab}) may suggest a finely tuned
pressure-confined source.  This makes 3C~346 different from most other
X-ray detected FRII \citep{hard00b} and FRI radio galaxies
\citep[e.g.][]{worr00}, which need an additional source of internal
pressure in order to balance the external gas pressure.

\section{The Unresolved X-ray Emission}\label{sec-agn}

Unresolved emission dominates the X-radiation at both ROSAT and ASCA
energies.  The measurements are summarized in Table~\ref{agn-tab}.
The fact that the unresolved emission is not heavily absorbed (with an
upper limit of $\sim 2\times 10^{21}$ cm$^{-2}$) means either that it
is nonthermal and related to the radio structures on VLBI to arcsec
scales, or that the central AGN regions are not hidden by obscuration.
The low absorption and X-ray variability support conclusions based on
VLBI data \citep{cotton95} that the source is a foreshortened FRII,
and so it should be more closely related to the quasar than
radio-galaxy class.

The absence of broad H$\beta$ is then unusual.  Explaining this by an
extinction of $A_v = 8$~mag \citep{dey94}, and adopting the gas to
dust ratio applicable to our own galaxy, implies that the broad-line
regions, and so too presumably the AGN, lie beyond gas which provides
a line of sight Hydrogen column density of $N_H = 1.4 \times 10^{22}$
cm$^{-2}$ \citep{burst78}.  We have placed an upper limit on an X-ray
component this heavily obscured by fitting the ASCA data to thermal
emission (fixed as before) together with two power-law components, one
of which is absorbed by $N_H = 1.4 \times 10^{22}$ cm$^{-2}$ and one
of which suffers no intrinsic absorption in the source.  These fits
were run for two adopted spectral indices for the absorbed power law:
$\alpha = 0.7$ and $\alpha = 1$ (photon index of 1.7 and 2.0).  In
neither case were the fitted spectral indices and normalizations for
the unabsorbed emission much changed from Figure~\ref{ascaconts},
although the uncertainties in spectral index increased as expected,
particularly to allow steeper slopes.  Although on the basis of an
F-test there is no justification for including the extra (absorbed)
component, the 90\% upper limits on its intensity are quite high: in
units of photons cm$^{-2}$ s$^{-1}$ keV$^{-1}$ at 1~keV, the
normalizations are $1.25 \times 10^{-4}$ and $1.9 \times 10^{-4}$ for
slopes of $\alpha = 0.7$ and 1.0, respectively.  These values
correspond to an upper limit on the pre-absorption 2-10 keV X-ray
luminosity of $5.6 \times 10^{43}$ ergs s$^{-1}$, or about 50\% as
luminous as the detected X-rays.  More sensitive high-energy X-ray
measurements are required to investigate whether such an absorbed
central X-ray component is present in 3C~346.

It seems reasonable that the unabsorbed unresolved X-ray emission
comes primarily from the inner regions of the radio jet, within an
arcsec of the core \citep{worr97}.  Indeed, the ratio of 3C~346's
radio-core and unresolved-X-ray flux densities are in agreement with
those for other 3CRR radio galaxies and quasars, from which we have
made a statistical argument that each source has a beamed nuclear soft
X-ray component directly related to the radio core \citep{hard99}.  In
common with other 3CRR radio galaxies and quasars, the core radio
spectrum is flat, with VLA flux densities of 220~mJy at 5~GHz and
243~mJy at 15 GHz, and a 5~GHz VLBI measurement of 165~mJy
\citep{vanb92, giov90}.  An optical core component of $37.5~\mu$Jy at
7000~\AA\ has been separated from the galaxy in HST data by
\citet{chiab99}.  Combined with our X-ray measurements, these results
imply interpolated two-point spectral indices of $\alpha_{\rm ro} =
0.76$, $\alpha_{\rm ox} = 0.86$, and $\alpha_{\rm rx} = 0.8$, and
place 3C~346 within the color range of low-power FRI radio galaxies
\citep{hard00a}. Radio-loud quasars from 3CRR typically have steeper
values of $\alpha_{\rm ox}$, in the range 1.0 to 1.6 \citep{wilkes94}.
The spectral energy distribution therefore argues that the viewing
angle to 3C~346 is not too much smaller than $30^\circ$, placing it on
the boundary between a radio galaxy and quasar.

It is likely that a significant fraction of the X-ray emission which
is unresolved to ROSAT and ASCA actually comes from the radio and
optical knots, and in particular the brightest of these, knot C, $3''$
from the core.  The radio spectral index in the knot between 1.6 and
15 GHz \citep{spence91, dey94, vanb92} is roughly $\alpha = 0.5$,
consistent with synchrotron emission from an electron population which
is energized by first-order Fermi acceleration at strong shocks.  The
optical flux-density \citep{devries97} is below an $\alpha=0.5$
extrapolation from 15 GHz, implying an energy-loss break in the
electron spectrum.  Modelling the knot as a sphere of radius 165~pc,
based on the 15~GHz measurements \citep{vanb92}, we find a good fit to
the radio and optical data for an equipartition magnetic field of
$\sim 650~\mu$Gauss and an electron spectral index which breaks by
unity at $10^{10}$~eV.  If, as is likely, the electron spectrum
extends to $10^{13}$~eV, we expect significant X-ray synchrotron
emission from the knot, amounting to roughly 10\% of the total
unresolved emission.  This should be easily detectable with {\it
Chandra\/}, which has the spatial resolution to separate the core and
jet.  {\it Chandra\/} observations will also make a strong test of the
equipartition assumption which has been brought into question by the
results for the jet in quasar PKS~0637-752 \citep{chartas00,
schwartz00}. If the X-ray emission in 3C~346's knot is above the
synchrotron prediction, then we will be forced to infer a
Compton-scattering origin, either from the synchrotron self-Compton
process - which is most effective if the source is out of
equipartition in the sense of weak magnetic field strength -- or from
Compton scattering of external photons.  If this is the case, then an
even higher fraction of the unresolved X-rays might originate in the
jet.

\section{Conclusions}\label{sec-conclusions}

\begin{figure}
\epsscale{0.8}
\plotone{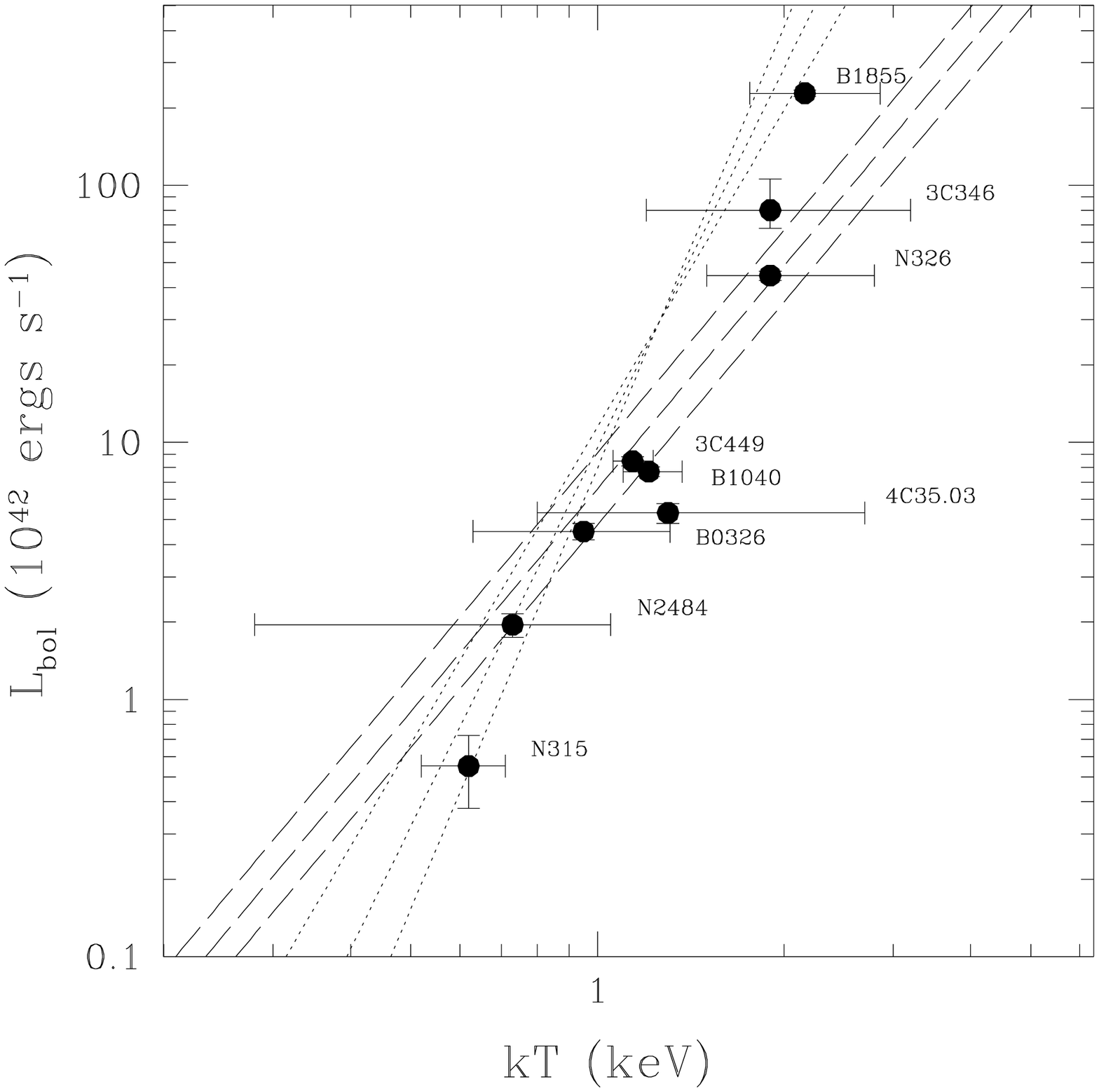}
\caption{Luminosity and temperature of 3C~346's X-ray emitting
atmosphere compared with results for the atmospheres of FR~I radio
galaxies from \citet{worr00}.  The dashed lines show the
temperature-luminosity relation (and errors) for more luminous
clusters ($\sim 10^{44} - 10^{46}$ ergs s$^{-1}$) from \citet{arn99}.
The dotted lines are the same for optically-selected X-ray bright
groups from \citet{helpond00}, where the steeper correlation is argued
to be in support of gas preheating which inhibits its collapse into the
shallow potential wells of poor systems.
\label{templum}}
\end{figure}

ROSAT and ASCA measurements of the nearby powerful radio source 3C~346
find extended emission, consistent with a cluster atmosphere, and
unabsorbed unresolved emission, consistent with non-thermal radiation
from radio structures on VLBI to arcsecond scale sizes.  We measure a
temperature of $1.9^{+1.3}_{-0.7}$ keV for the extended emission,
making this the second low-redshift ($z < 0.2$) powerful radio galaxy,
after Cygnus~A, with a measured cluster temperature.  The
temperature and luminosity of the cluster gas are consistent with the
correlation found for the atmospheres around nearby, less powerful,
radio galaxies (Fig.~\ref{templum}).  The cooling time for the cluster
gas is too long for a significant cooling flow to have become
established.  The radio lobes of 3C~346 are roughly in pressure
balance with the external medium under the assumptions that the energy
densities in the magnetic field and radiating particles balance, and
that a source of excess pressure in the radio lobes, commonly invoked
in other radio galaxies, is absent here.

3C~346's orientation to the line of sight is uncertain.  It was
originally classified as being an intrinsically small member of the
CSS class, but radio measurements have since been used to infer a
jet-angle to the line of sight of $\theta < 32^\circ$. The evidence
that the unresolved X-ray flux is not only unabsorbed but has varied
by $32\pm 13$\% over 18 months makes it likely that a significant
fraction of this emission is from deep in the radio jets, in
sub-arcsec regions and influenced by relativistic beaming, which would
support a small angle to the line of sight.  However, the
radio/optical/X-ray colors are not those of a quasar, rather they are
similar to those of other nearby radio galaxies, and so we support the
idea that this source is at an orientation intermediate between
quasars and radio galaxies.

3C~346 is one of a relatively small number of radio galaxies where
optical emission has been detected from the radio jet.  We predict
that the brightest knot in the jet is detectable with {\it Chandra\/},
and observations will test the equipartition assumptions which have
been challenged recently by the {\it Chandra\/} observations of
PKS~0637-752.  {\it Chandra\/} observations are also required to probe
the possible presence of an additional heavily absorbed X-ray emission
component of $< 5.6 \times 10^{43}$ ergs s$^{-1}$ (2-10~keV), where
the absorption is due to the gas suggested to be obscuring the
broad-line emission regions in this source.

\acknowledgments

We acknowledge support from NASA grant NAG~5-2961.

\clearpage

\clearpage

\clearpage

\clearpage

\clearpage

\clearpage

\clearpage

\clearpage

\begin{deluxetable}{llllllll}
\tablecaption{X-ray Observations\label{obs-tab}}
\tablecolumns{7}
\tablewidth{0pc}
\tablehead{
\colhead{Source} 
        & \colhead{$z$}
        & \colhead{N$_{\rm H}$ $^{\rm a}$}
        & \colhead{Date}
        & \colhead{Mission}
        & \colhead{Instrument}
        & \colhead{Energy}
        & \colhead{Exposure} \\
\colhead{} 
        & \colhead{}
        & \colhead{(cm$^{-2}$)}
        & \colhead{}
        & \colhead{}
        & \colhead{}
        & \colhead{Band (keV)}
        & \colhead{(ks)} \\
}
\startdata
3C 346 & 0.161 & $5.47 \times 10^{20}$& 1993 Aug 14-17 & ROSAT & PSPC
& 0.2--2.5 & 16.9 \\
 & & & 1995 Feb 17-18 & ASCA & SIS, GIS & 0.4--10 & 21.6 \\
\tablenotetext{a}{From \citet{stark92}}
\enddata
\end{deluxetable}

\begin{deluxetable}{ll}
\tablecaption{Parameters for the Cluster X-ray Emission around 3C~346. 
\label{cluster-tab}}
\tablewidth{0pt}
\tablehead{
\colhead{Parameter}     & \colhead{Value} }
\startdata
ROSAT counts (0.2-1.9 keV), $\theta < 3'$ & 
$371^{+104}_{-46}$ \\
$\beta$ $^{\rm a}$& $0.9^{+3.1}_{-0.44}$  \\
$\theta_{\rm cx}$ $^{\rm a}$& $78^{+167}_{-55}$ arcsec \\
$kT$ & $1.9^{+1.3}_{-0.7}$~keV  \\
Density: $\theta = 0$, $\theta = 10''$ & 
  $1.6^{+0.8}_{-0.4} \times 10^{-3}$, 
  $1.5^{+0.75}_{-0.3} \times 10^{-3}$ cm$^{-3}$  \\
Pressure: $\theta = 0$, $\theta = 10''$ & 
  $1.1^{+1.0}_{-0.5} \times 10^{-11}$,
  $1.05^{+0.9}_{-0.4} \times 10^{-11}$ dynes cm$^{-2}$ \\
Cooling time: $\theta = 0$, $\theta = 10''$ & 
  $2.6^{+1.2}_{-1.0} \times 10^{10}$,
  $2.7^{+1.2}_{-1.0} \times 10^{10}$ years \\
$L_{0.2 - 2.5~\rm keV}$ & $5.3^{+1.6}_{-0.8} \times 10^{43}$ ergs s$^{-1}$ \\
$L_{0.5 - 3~\rm keV}$ & $4.4^{+1.4}_{-0.6} \times 10^{43}$ ergs s$^{-1}$ \\
$L_{2 - 10~\rm keV}$ (extrapolated)& $2^{+1}_{-0.6} \times 10^{43}$ ergs s$^{-1}$ \\
$L_{\rm Bol}$ (extrapolated)& $8^{+2.6}_{-1.2} \times 10^{43}$ ergs s$^{-1}$ \\
\tablenotetext{a}{$\beta$ and $\theta_{\rm cx}$ are highly
correlated and errors are for two interesting parameters.  
See Figure~\ref{beta346}}
\enddata
\tablecomments{All measurements are from the ROSAT PSPC observation.
Errors are $1\sigma$.
Multiply pressure values by 0.1 to give in units of 
N m$^{-2}$ (Pascals).}
\end{deluxetable}

\begin{deluxetable}{ll}
\tablecaption{Parameters for the Active Galaxy X-ray Emission in 3C~346. 
\label{agn-tab}}
\tablewidth{0pt}
\tablehead{
\colhead{Parameter}     & \colhead{Value} }
\startdata
ROSAT counts (0.2-1.9 keV) & $736^{+21}_{-34}$ \\
ROSAT Power-law energy index, $\alpha_x$ & $0.69^{+0.16}_{-0.14}$ \\
ASCA Power-law energy index, $\alpha_x$ & $0.73^{+0.17}_{-0.23}$ \\
ROSAT PSPC $f_{1~\rm keV}$  & $0.162^{+0.004}_{-0.008} \mu$Jy \\
ASCA $f_{1~\rm keV}$  & $0.11^{+0.02}_{-0.03} \mu$Jy \\
ROSAT $L_{0.2 - 2.5~\rm keV}$ & $1.14^{+0.03}_{-0.05}  \times 10^{44}$ 
ergs s$^{-1}$ \\
ASCA $L_{0.2 - 2.5~\rm keV}$ (extrapolated) & 
$7.8^{+1.1}_{-1.7}  \times 10^{43}$ ergs s$^{-1}$ \\
ROSAT $L_{0.5 - 3~\rm keV}$ & $9.5^{+0.3}_{-0.4}  \times 10^{43}$ 
ergs s$^{-1}$ \\
ASCA $L_{0.5 - 3~\rm keV}$ &
$6.4^{+0.8}_{-1.4}  \times 10^{43}$ ergs s$^{-1}$ \\
ROSAT $L_{2 - 10~\rm keV}$ (extrapolated)& $1.28^{+0.04}_{-0.07} 
 \times 10^{44}$ ergs s$^{-1}$ \\
ASCA $L_{2 - 10~\rm keV}$ & 
$8.1^{+1.2}_{-1.8}  \times 10^{43}$ ergs s$^{-1}$ \\
\enddata
\tablecomments{Errors are $1\sigma$.}
\end{deluxetable}


\clearpage

\begin{thebibliography}{}

\bibitem[Arnaud \& Evrard(1999)]{arn99}
Arnaud, M. \& Evrard, A.E.~1999, \mnras, 305, 631

\bibitem[Arnaud et al.(1992)]{arnaud92}
Arnaud, M., Rothenflug, R., Boulade, O., Vigroux, L. \& Vangioni-Flam,
E.~1992, \aap, 254, 49

\bibitem[Barthel(1989)]{bart89}
Barthel, P.D.~1989, \apj, 336, 606

\bibitem[Baum et al.(1988)]{baum88}
Baum, S.A., Heckman, T.M. \& van Breugel, W.J.M.~1988, \apjs, 68, 643

\bibitem[Birkinshaw(1994)]{birk94}Birkinshaw, M.~1994, in Astronomical
Data Analysis Software and Systems III, ASP Conference Series Volume
61, eds. D.R. Crabtree, R.J. Hanisch \& J. Barnes, 249.

\bibitem[Burstein \& Heiles (1978)]{burst78}
Burstein, D. \& Heiles, C.~1978, \apj, 225, 40


\bibitem[Chartas et al.(2000)]{chartas00}
Chartas, G. et al.~2000, \apj, 542, 655

\bibitem[Chiaberge et al.(1999)]{chiab99}
Chiaberge, M., Capetti, A. \& Celotti, A.~1999, \aap, 349, 77

\bibitem[Cotton et al.(1995)]{cotton95}
Cotton, W.D., Feretti, L., Giovannini, G., Venturi, T., Lara, L.,
Marcaide, J. \& Wehrle, A.E.~1995, \apj, 452, 605

\bibitem[de Koff et al.(1996)]{dekoff96}
de Koff, S, Baum, S.A., Sparks, W.B., Biretta, J., Golombek, D., 
Macchetto, F., McCarthy, P. \& Miley, G.K.~1996, \apjs, 107, 621

\bibitem[de Vries et al.(1997)]{devries97}
de Vries, W.H., O'Dea, C.P., Baum, S.A., Sparks, W.B., Biretta, J., de
Koff, S., Golombek, D., Lehnert, M.D., Macchetto, F., McCarthy, P. \&
Miley, G.K.~1997, \apjs, 110, 191

\bibitem[Dey \& van Breugel(1994)]{dey94}
Dey, A. \& van Breugel, W.J.M.~1994, \aj, 107, 1977

\bibitem[Fabbiano et al.(1992)]{fab92}
Fabbiano, G., Kim, D.-W. \& Trinchieri, G.~1992, 
\apjs, 80, 513

\bibitem[Fabbiano et al.(1984)]{fab84}
Fabbiano, G., Miller, L., Trinchieri, G., Longair, M.~\& Elvis, M.
1984, \apj, 277, 115

\bibitem[Fanaroff \& Riley(1974)]{fr74}
Fanaroff, B.L. \& Riley, J.M.~1974, \mnras, 167, 31P

\bibitem[Fanti et al.(1985)]{fanti85}
Fanti, C., Fanti, R., Parma, P., Schilizzi, R.T. \& van Breugel,
W.J.M.~1985, \aap, 143, 292

\bibitem[Giovannini et al.(1990)]{giov90}Giovannini, G., Feretti,
L. \& Comoretto, G.~1990, \apj, 358, 159

\bibitem[Hardcastle \& Worrall(1999)]{hard99}Hardcastle, M.J. \&
Worrall, D.M.~1999, \mnras, 309, 969

\bibitem[Hardcastle \& Worrall(2000a)]{hard00a}Hardcastle, M.J. \&
Worrall, D.M.~2000a, \mnras, 314, 359

\bibitem[Hardcastle \& Worrall(2000b)]{hard00b}Hardcastle, M.J. \&
Worrall, D.M.~2000b, \mnras, in press

\bibitem[Helsdon \& Ponman(2000)]{helpond00}
Helsdon, S.F. \& Ponman, T.J.~2000, \mnras, 315, 356


\bibitem[Laing et al.(1983)]{lrl83}Laing, R.A., Riley, J.M. \&
Longair, M.S.~1983, \mnras, 204, 151


\bibitem[Preibisch(1997)]{prei97}Preibisch, T.~1997,
\aap, 320, 525

\bibitem[Sambruna et al.(1999)]{sam99}Sambruna, R.M., Eracleous, M. \&
Mushotzky, R.F.~1999, \apj, 526, 60

\bibitem[Sarazin(1986)]{sar86}Sarazin, C.L.~1986, Rev. Mod. Phys., 58, 1

\bibitem[Sciortino et al.(1995)]{scio95}Sciortino, S., Favata, F. \&
Micela, G.~1995, \aap, 296, 370

\bibitem[Schwartz et al.(2000)]{schwartz00}
Schwartz, D.A. et al.~2000, \apj, 540, L69.

\bibitem[Spencer et al.(1991)]{spence91}Spencer, R.E., Schilizzi,
R.T., Fanti, C., Fanti, R., Parma, P., van Breugel, W.J.M., Venturi,
T., Muxlow, T.W.B. \& Rendong, N.~1991, \mnras, 250, 225

\bibitem[Stark et al.(1992)]{stark92}Stark, A.A., Gammie, C.F.,
Wilson, R.W., Bally, J., Linke, R.A., Heiles, C. \& Hurwitz, M.~1992,
\apjs, 79, 77

\bibitem[van Breugel et al.(1992)]{vanb92}van Breugel, W.J.M., Fanti,
C., Fanti, R., Stanghellini, C., Schilizzi, R.T. \& Spencer, R.E.~1992,
\aap, 256, 56

\bibitem[Wilkes et al.(1994)]{wilkes94}Wilkes, B.J., Tananbaum, H.,
Worrall, D.M., Avni, Y., Oey, M.S. \& Flanagan, J.~1994, \apjs, 92, 53

\bibitem[Worrall(1997)]{worr97}Worrall D.M.~1997, in {\it Relativistic
Jets in AGNs}, ed. M.~Ostrowski, M.~Sikora, G.~Madjeski \& M.~Begelman
(Astronomical Observatory of the Jagiellonian University, Krakow), 20
(astro-ph/9709165) 

\bibitem[Worrall \& Birkinshaw(1994)]{worr94}
Worrall, D.M. \& Birkinshaw, M.~1994, \apj, 427, 134

\bibitem[Worrall \& Birkinshaw(2000)]{worr00}
Worrall, D.M. \& Birkinshaw, M.~2000, \apj, 530, 719

\bibitem[Zirbel(1997)]{zirb97}
Zirbel, E.L.~1997, \apj, 476, 489


\end{thebibliography}
\end{document}